\documentclass{article}

\usepackage{amsmath,amsfonts,bm}

\def\eqref#1{equation~\ref{#1}}

\def\1{\bm{1}}

\def\mG{{\bm{G}}}
\def\mH{{\bm{H}}}
\def\mI{{\bm{I}}}

\def\mL{{\bm{L}}}

\def\mP{{\bm{P}}}
\def\mQ{{\bm{Q}}}
\def\mR{{\bm{R}}}

\def\mW{{\bm{W}}}

\DeclareMathAlphabet{\mathsfit}{\encodingdefault}{\sfdefault}{m}{sl}
\SetMathAlphabet{\mathsfit}{bold}{\encodingdefault}{\sfdefault}{bx}{n}

\newcommand{\R}{\mathbb{R}}

\usepackage{microtype}
\usepackage{booktabs} 

\usepackage{hyperref}

\usepackage[accepted]{icml2023}

\usepackage{amsmath}
\usepackage{amssymb}
\usepackage{mathtools}
\usepackage{amsthm}
\usepackage{times}
\usepackage{url}
\usepackage[pdftex]{graphicx}
\usepackage{algorithmic}
\usepackage{algorithm}
\usepackage{bm}
\usepackage{setspace}
\usepackage{booktabs}
\usepackage{caption}
\usepackage{subcaption}
\usepackage{multirow}

\usepackage[capitalize,noabbrev]{cleveref}

\theoremstyle{plain}

\theoremstyle{definition}

\theoremstyle{remark}

\usepackage[textsize=tiny]{todonotes}

\icmltitlerunning{Do You Remember? Overcoming Catastrophic Forgetting for Fake Audio Detection}

\begin{document}

\twocolumn[
\icmltitle{Do You Remember? Overcoming Catastrophic Forgetting for Fake Audio Detection}




\begin{icmlauthorlist}
\icmlauthor{Xiaohui Zhang}{nlpr,bjtu}
\icmlauthor{Jiangyan Yi}{nlpr}
\icmlauthor{Jianhua Tao}{tsu}
\icmlauthor{Chenglong Wang}{nlpr,stc}
\icmlauthor{Chuyuan Zhang}{nlpr}
\end{icmlauthorlist}

\icmlaffiliation{bjtu}{School of Computer and Information Technology, University of Beijing Jiaotong, Beijing, China}
\icmlaffiliation{nlpr}{State Key Laboratory of Multimodal Artificial Intelligence Systems, Institute of Automation, Chinese Academy of Sciences, Beijing, China}
\icmlaffiliation{tsu}{Department of Automation, Tsinghua University, Beijing, China}
\icmlaffiliation{stc}{University of Science and Technology of China, Beijing, China}
\icmlcorrespondingauthor{Jiangyan Yi}{jiangyan.yi@nlpr.ia.ac.cn}

\icmlkeywords{Machine Learning, ICML}

\vskip 0.3in
]



\printAffiliationsAndNotice{} 

\begin{abstract}
Current fake audio detection algorithms have achieved promising performances on most datasets. However, their performance may be significantly degraded when dealing with audio of a different dataset. The orthogonal weight modification to overcome catastrophic forgetting does not consider the similarity of genuine audio across different datasets. To overcome this limitation, we propose a continual learning algorithm for fake audio detection to overcome catastrophic forgetting, called Regularized Adaptive Weight Modification (RAWM). When fine-tuning a detection network, our approach adaptively computes the direction of weight modification according to the ratio of genuine utterances and fake utterances. The adaptive modification direction ensures the network can effectively detect fake audio on the new dataset while preserving its knowledge of old model, thus mitigating catastrophic forgetting. In addition, genuine audio collected from quite different acoustic conditions may skew their feature distribution, so we introduce a regularization constraint to force the network to remember the old distribution in this regard. Our method can easily be generalized to related fields, like speech emotion recognition. We also evaluate our approach across multiple datasets and obtain a significant performance improvement on cross-dataset experiments.
\end{abstract}

\section{Introduction}
With the development of speech synthesis and voice conversion technology \citep{Wang2018StyleTU, Wang2021ProsodyAV}, the models can generate human-like speech, which makes it difficult for most people to distinguish the generated audio from the real one. Although this technology has brought great convenience to human life, it has also brought great safety hazards to the country and society. Therefore, fake audio detection has attracted increasing attention in recent years.
A series of challenges have been organized to detect fake audio,  
\begin{figure*}
\begin{tiny}
\begin{center}
\begin{subfigure}[ht]{0.4\linewidth}
\begin{center}
\includegraphics[width=0.7\linewidth]{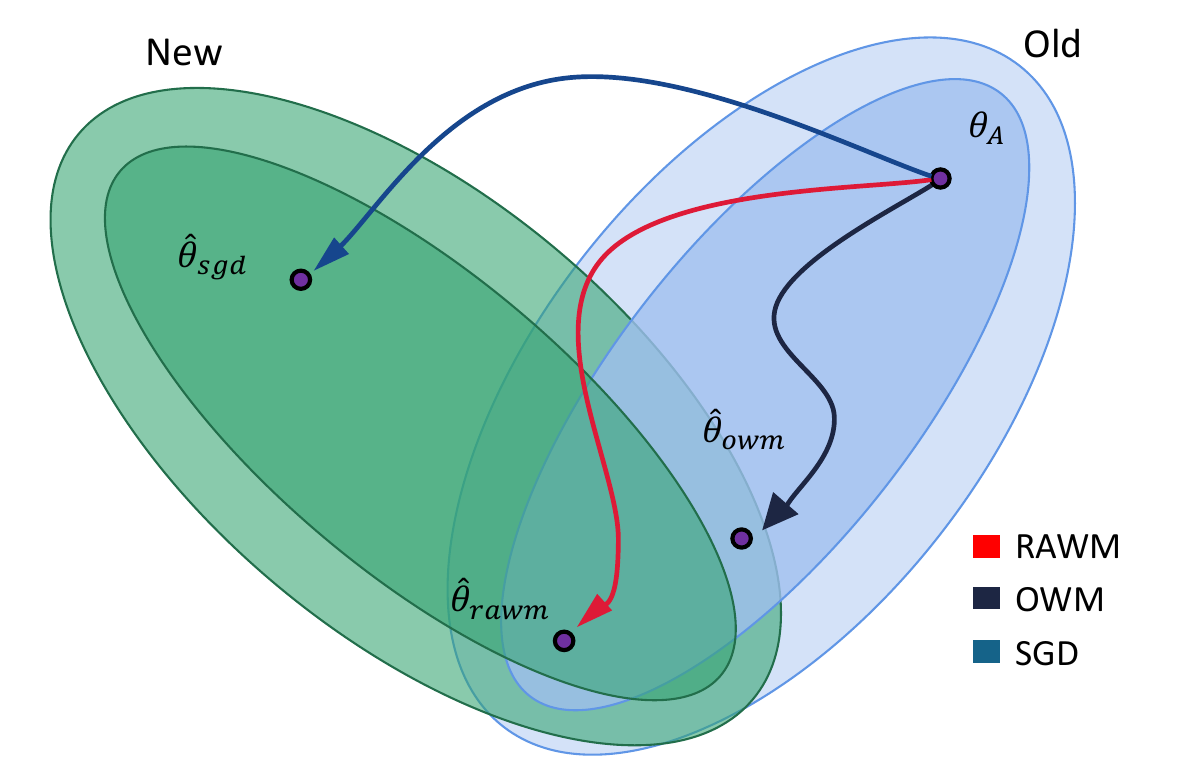}
\caption{}
\label{opticompare}
\end{center}
\end{subfigure}
\begin{subfigure}[ht]{0.55\linewidth}
\begin{center}
\includegraphics[width=0.7\linewidth]{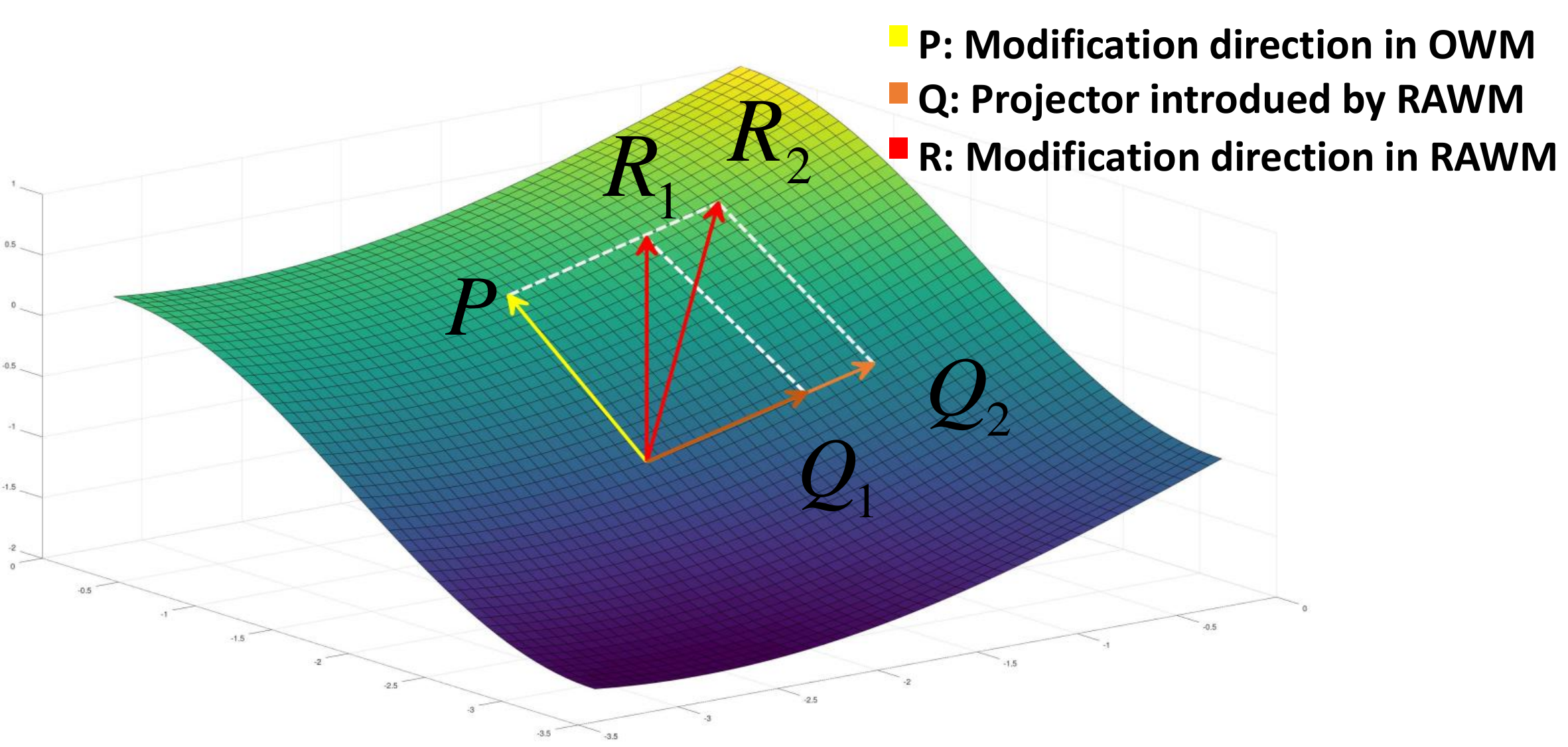} 
\caption{}
\label{rawmfb}
\end{center}
\end{subfigure}			
\caption{Schematic of SGD, OWM, and RAWM. \textbf{(a)}, With RAWM, the optimization process searches for configurations that lead to great performance on both old (blue area) and new (green area) datasets. The center parts of the two areas represent better recognition performance than the other, and can be regarded as subspaces of the area mentioned by the OWM. A successful optimized configuration $\hat{\theta}_{rawm}$ stops inside the overlapping subspace. However, the configuration $\hat{\theta}_{sgd}$ obtained by SGD is optimized without considering forgetting, and the configuration $\hat{\theta}_{owm}$ obtained by orthogonal weight modification can reach the overlapping area but not the overlapping subspace. \textbf{(b)}, the RAWM adaptively modifies weight direction by introducing a projector that is orthogonal to the projector $\mP$ proposed by OWM.}
\label{awm_fig}
\end{center}
\vspace{-2em}
\end{tiny}
\end{figure*}
such as the ASVspoof challenge \citep{wu2015asvspoof, kinnunen2017asvspoof, todisco2019asvspoof, yamagishi2021asvspoof} and the Audio Deep Synthesis Detection (ADD) challenge \citep{yi2022add}. In these competitions, deep neural networks have achieved great success. Currently, large-scale pre-trained models have gradually been applied to fake audio detection and achieved state-of-the-art results on several public fake audio detection datasets \citep{tak2022automatic, martin2022vicomtech, lv2022fake, wang2021investigating}. Although fake audio detection achieves promising performance, it may be significantly degraded when dealing with audio of another dataset. The diversity of audio proposes a significant challenge to fake audio detection across datasets \citep{zhang2021empirical, zhang2021one}.\par
Some approaches have been proposed to improve detection performance across datasets. An ensemble learning method is proposed to improve the detection ability of the model for unseen audio \citep{MonteiroAF2020} and a dual-adversarial domain adaptive network (DDAN) is designed to learn more generalized features for different datasets \citep{WangD0Q020}. Both methods require some audio from the old dataset, but in some practical situations, it is almost impossible to obtain them. For instance, a pre-trained model proposed by a company has been released to the public. It is unfeasible for the public to fine-tune it using the data belonging to the original company. In addition, a data augmentation method is proposed to extract more robust features for detection across datasets \citep{zhang2021empirical}, which is only suitable for the datasets with similar feature distribution. In continual learning, a method called Detecting Fake Without Forgetting (DFWF) is proposed for fake audio detection \citep{ma2021continual}. 
Although the above methods are effective, they still have some limitations, like the acquisition of old data in the ensemble learning method and the DDAN and deteriorating learning performance in the DFWF. This paper, however, aims to overcome catastrophic forgetting while exerting a positive influence on acquiring new knowledge without any previous samples.\par
Most fake audio detection datasets are under clean conditions, where the genuine audio has a more similar feature distribution than the fake audio \citep{ma2021continual}. A few datasets, however, are collected under different acoustic conditions \citep{muller2022does}, which makes a difference in their feature distributions of genuine audio \citep{DBLP:journals/corr/abs-2207-12308}. If we modify the model weights as the orthogonal weight modification (OWM) method \citep{zeng2019continual} which introduces a new weight direction orthogonal to all old data, most genuine audio with similar feature distribution across datasets can not be trained effectively. The reason is that new genuine audio is supposed by the OWM to damage learned knowledge, so it modifies new weight direction orthogonal to the old one regardless the new and old genuine audio have similar feature distribution and they can be seen as a whole from the same dataset. Based on the above inference, it is more effective for genuine audio on new datasets to be trained with a direction close to the previous one, rather than orthogonal to it. 
To address these issues, we propose a continual learning approach, named Regularized Adaptive Weight Modification (RAWM). 
In our method, if the proportion of fake audio is larger, the modified direction is closer to the orthogonal projector of the subspace spanned by all old input; if the proportion of genuine audio is larger, the modification is closer to the old input subspace. However, old and new datasets are collected from different acoustic conditions in some cases, where genuine audio may have quite different feature distributions. We address this issue by introducing a regularization constraint. This constraint forces the model to remember the old feature distribution. In addition, compared with the experience-replay-based method, RAWM does not require old data, which makes it suitable for most situations. The optimization process of RAWM is compared with that of the Stochastic Gradient Descent search (SGD) and OWM in Figure \ref{opticompare}.\par
\textbf{Contributions:} We propose a regularized adaptive weight modification algorithm to overcome catastrophic forgetting. There are two essential steps in our method: adaptive weight modification (AWM) and regularization. The former AWM is proposed for continual learning in most situations where genuine audio has similar feature distribution and the latter regularization is introduced to ease the problem that genuine audio may have different feature distribution in a few cases.
Although our method is inspired by the feature distribution similarity in fake audio detection, it can also be used in other related tasks, such as speech emotion recognition.
The experimental results show that our proposed method outperforms several continual learning methods in acquiring new knowledge and overcoming forgetting, including Elastic Weight Consolidation (EWC) \citep{kirkpatrick2017overcoming}, Learning without Forgetting (LwF) \citep{li2017learning}, OWM, and DFWF.
The code of our method has been released in \href{https://github.com/Cecile-hi/Regularized-Adaptive-Weight-Modification.git}{Regularized Adaptive Weight Modification}.
\section{Related Work}
\label{related}
In continual learning, overcoming catastrophic forgetting methods can be divided into the following categories. The regularization methods perform a regularization on the objection function or regulate important weights that are essential for previous tasks \citep{kinnunen2017asvspoof, zenke2017continual, aljundi2018memory, aljundi2019task, mallya2018packnet, serra2018overcoming}. The dynamic architecture methods reserve their previous knowledge by introducing additional layers or nodes and grow model architecture \citep{rusu2016progressive, schwarz2018progress}; \citep{yoon2017lifelong}. The memory-based methods remember their previous data to prevent gradient updates from damage on their learned knowledge. \citep{lopez2017gradient, castro2018end, wu2019large, lee2019overcoming}. 
The natural gradient descent methods approximate the Fisher information matrix in EWC using the generalized Gauss-Newton method to fast gradient descent \citep{tseran2018natural, chen2019facilitating}.
\par
Although the mainstream continual learning methods, such as the EWC, LwF and OWM, have achieved great success in many fields including image classification \citep{zeng2019continual, kirkpatrick2017overcoming}, object detection \citep{perez2020incremental}, semantic segmentation \citep{cermelli2020modeling}, lifelong language learning \citep{de2019episodic} and sentence representation \citep{liu2019continual}. However, the approximation of regularization methods will produce error accumulation in continual learning \citep{DBLP:conf/icml/ZenkePG17, DBLP:journals/corr/abs-1712-03847, ma2021continual}. In contrast, our proposed method only needs the current inputs, which leads to a better performance than others in error accumulation. Besides, we relax the regularized constraint in the DFWF and introduce a direction modification to solve the deteriorating learning performance problem. 
\section{Background}
\label{Bbackground}
\subsection{Orthogonal Weight Modification}
\label{owm}
The OWM algorithm overcomes catastrophic forgetting by modifying the direction of weights on the new task. The modified direction $\mP$, which is a square matrix, is orthogonal to the subspace spanned by all inputs of the previous task. The orthogonal projector is constructed by an iterative method similar to the Recursive Least Square (RLS) algorithm \citep{shah1992optimal}, which hardly requires any previous samples.\par
We consider a feed-forward network consisting of $L+1$ layers, indexed by $l=0,1,\cdots,L$ with the same activation function $g(\cdot)$. The $\mathbf{\overline{x}}_l(i,j)\in \R^{s}$ represents the output of the $l$th layer in response to the mean of the $i$th batch inputs on $j$th dataset, and the $\mathbf{\overline{x}}_l(i,j)^T$ is the transpose matrix of the $\mathbf{\overline{x}}_l(i,j)$. The modified direction $\mP$ can be calculated as:
	\begin{equation}
	\small
	\begin{split}
	\mP_l(i,j)&=\mP_l(i{-1},j)-\mathbf{k}_l(i,j)\mathbf{\overline{x}}_{l{-1}}(i,j)^T \mP_l(i{-1},j)
	\\   \mathbf{k}_l(i,j)&=\frac{\mP_l(i{-1},j)\mathbf{\overline{x}}_{l{-1}}(i,j)}{\alpha+\mathbf{\overline{x}}_{l{-1}}(i,j)^T \mP_l(i{-1},j)\mathbf{\overline{x}}_{l{-1}}(i,j)}
	\end{split}
	\label{compute_P}
	\end{equation}
where $\alpha$ is a hyperparameter decaying with the number of tasks. 
\subsection{Learning without Forgetting}
\label{lwf}
The LwF algorithm is inspired by the idea of model distillation, where old knowledge is viewed as a penalty term to regulate the new model representation similar to the old. The model trained on old datasets is replicated into two models with the same parameters. The two models are named teacher and student models in the LwF. In process of training on new datasets, the parameters of the teacher model are frozen to produce its features as "soft labels". The student model is trained by the loss function as:
\begin{equation}
\small
\mL_{lwf}=\lambda_0\mL_{old}(\mathrm {y}_o, \hat{\mathrm{ y}}_o)+\mL_{new}(\mathrm{ y}_n,\hat{\mathrm{ y}}_n)
\label{backformula_lwf}
\end{equation}
where $\lambda_0$ is a ratio coefficient representing the importance of learned knowledge; $\mathrm{y_o}$ is the "soft label" produced by the teacher model and $\mathrm{y_n}$ is the ground truth of new data; Both $\hat{\mathrm{y}}_o$ and $\hat{\mathrm{y}}_n$ are the softmax output of the student model. Both $\mL_{old}$ and $\mL_{new}$ are cross-entropy loss. The former $\mL_{old}$ regulates the output probabilities $\hat{\mathrm{y}}_o$ to be close to the recorded output $\mathrm{y_o}$ from the teacher model and the latter $\mL_{new}$ encourages predictions $\hat{\mathrm{y}}_n$ to be consistent with the ground truth $\mathrm{y_n}$.
\section{Proposed Method}
On most fake audio detection datasets, under the same acoustic conditions, feature distributions of genuine audio are relatively more concentrated than the fake, which means the feature distribution of genuine audio has a smaller variance than that of fake audio \citep{ma2021continual, DBLP:journals/corr/xinrui}. Besides, there are also a few datasets whose genuine audio has quite different feature distributions from others \citep{DBLP:journals/corr/abs-2207-12308, muller2022does}. 
\par
Based on the observations, we propose a continual learning method, named Regularized Adaptive Weight Modification (RAWM), to overcome catastrophic forgetting.
There are two essential steps in our method: adaptive direction modification (AWM) and regularization. 
The AWM 
is proposed for most situations where genuine audio has similar feature distribution. As shown in Figure \ref{rawmfb}, by introducing an extra projector, which is a square matrix orthogonal to the projector proposed by the OWM, our method could adaptively modify weight direction closer to the previous inputs subspace. As for those genuine audio collected from quite different acoustic conditions, it is detrimental for learned knowledge to modify weight according to the rule we mentioned above, because their feature distribution is distinct from others. To address this issue, we introduce a regularization term to force the new distribution of inference to be similar to the old one. Our method does not require any replay of previous samples. In addition, our method is inspired by fake audio detection but it can easily be generalized to other related tasks. The reason is that most of them have one or more classes, like neutral emotion in speech emotion recognition (SER) \citep{9747417}, with relatively similar feature distribution between different datasets. We also take SER as an example to present how our method generalizes to other fields in Sec. \ref{rawm} and show the process of our algorithm in Algorithm \ref{alg:beam-search}.
\subsection{Adaptive Weight Modification}
\label{awm}
We start by introducing an adaptive modification of weight direction according to the ratio $\beta$ of classes with similar feature distribution between different datasets and others in batch data, which is essential for sequence training on multi-datasets. We first consider a feed-forward network like that described in Sec. \ref{owm}. Then, we introduce a square matrix $\mQ$ as a projector that is orthogonal to the $\mP$ proposed by the OWM algorithm. 
This orthogonal projector can be written as Eq \ref{compute_q}:
\begin{small}
	\begin{equation}
		\mQ=\beta[\mI\!-\!\mP(\mP^T \mP)^{-1} \mP^T]
		\label{compute_q}
		\end{equation}
	\end{small}
where the projector $\mP$, which is orthogonal to the subspace spanned by all previous inputs, can be calculated as Eq \ref{compute_P} and $\mI$ is an identity matrix.
The construction of the orthogonal projector $\mQ$ is mathematically sound \citep{haykin2002adaptive, ben2003generalized, Bengio+chapter2007}. 
To verify the modification direction according to the essential ratio $\beta$, we introduce the $\beta$ defined as:
\begin{equation}
\small
\beta=\dfrac{\sum\limits_{t=1}\limits^{b}N_t+1}{\sum\limits_{t=b+1}\limits^{b+c}N_t+1}
\label{betaall}
\end{equation}
in which the $N_t, t\in [1,b]$ represents the number of batch samples of $b$ classes with relatively similar feature distributions on old and new datasets, respectively; the $N_t, t\in [b+1,c]$ represents the number of batch samples of other $c$ classes.
By adding one to both the numerator and denominator, $\beta$ can be calculated when all the batch data belong to classes in the numerator. As illustrated in Eq \ref{compute_q}, the norm of projector $\mQ$ is proportional to the ratio $\beta$. Our approach defines the modified direction $\mR$ of weights as:
	\begin{equation}
	\small
	\mR\!=\!\mP_{norm}\!+\!\mathrm{m}\mQ_{norm}
	\label{compute_R}
	\end{equation}
	\begin{equation}
	\small
	\mP_{norm}\!=\!\dfrac{\mP}{|| \mP ||},\quad \mQ_{norm}\!=\!\dfrac{\mQ}{||\mI\!-\!\mP(\mP^T \mP)^{-1} \mP||}
	\end{equation}
where $\mathrm{m}$ is a constant to constrain the norm of projector $\mQ$ to prevent gradient explosion or gradient vanishing in the backward process; $\mP_{norm}$ and $\mQ_{norm}$ are identity matrices normalized by $\mP$ and $\mQ$, respectively. Normalization is to prevent the case that the change of $\beta$ has little effect on the modified direction because of the large norm gap between $\mP$ and $\mQ$. In the back-propagate (BP) process, the direction of network weights is modified as: 
\begin{small}
\begin{equation}
\begin{split}
\mW_l(i,j)\!=\!\mW_l(i{-1},j)\!+\!\mG &\quad j=1
\\
\mW_l(i,j)\!=\!\mW_l(i{-1},j)\!+\!\mR_l(j{-1})\mG &\quad j>1
\\
\quad\mG = \gamma(i,j) \Delta \mW^{BP}_l(i,j)
\end{split}
\label{bp_awm}
\end{equation}
\end{small}
where $\mW_l(i,j) \in \R^{s\times v}$ represents the connection weights between the $l$th layer and the $(l{+1})$th layer; $\gamma$ represents the learning rate of this network; $\Delta \mW_l^{BP}(i,j)$ represents the standard BP gradient; $\mR$
\begin{algorithm}[t] 
\setstretch{1.25}
\caption{Regularized Adaptive Weight Modification}\label{alg:beam-search}
\begin{algorithmic}[1]
\STATE {\bfseries Require:} Training data from different datasets, $\gamma$ (learning rate), $\mathrm{m}$ (constant hyperparameter), $T_{reg}$ (constant hyperparameter).
\FOR{every dataset \textit{j}}
\FOR {every batch \textit{i}}
\IF{$j =1$}
\STATE $\mW_l(i,j) = \mW_l(i{-1},j) + \gamma(i,j)\Delta \mW_l^{BP}(i,j)$
\ELSE
\STATE {$\mathbf{k}(i,j) = \dfrac{\mP_l(i{-1}) \mathbf{\overline{x}}_{l{-1}} (i,j)}{ \alpha + \mathbf{\overline{x}}_{l{-1}}(i,j)^T \mP_l(i{-1},j) \mathbf{\overline{x}}_{l{-1}} (i,j)}$}\\
\STATE {$\mP_l(i,j)\! =\! \mP_l(i{-1},j)\! -\! \mathbf{k}(i,j)\mathbf{\overline{x}}_{l{-1}}(i,j)^T \mP_l(i{-1},j)$} \\
\STATE {$\beta=\dfrac{\sum\limits_{t=1}\limits^{b}N_t+1}{\sum\limits_{t=b+1}\limits^{b+c}N_t+1}$} 
			\STATE {$\mQ=\beta[\mI-\mP(\mP^T \mP)^{-1} \mP^T]$}\\
			\STATE {$\mP_N=\dfrac{\mP}{|| \mP ||}$}\\
			\STATE {$\mQ_N=\dfrac{\mQ}{|| \mI\!-\!\mP(\mP^T \mP)^{-1} \mP ||}$}\\
			\STATE {$\mR=\mP_N+\mathrm{m}\mQ_N$}\\
			\STATE {$\hat{y}_o(i)= \frac{y_o(i)^{1/T_{reg}}}{\sum \mathbf{y}_o(i)^{1/T_{reg}}}$}\\
			\STATE {$\hat{y}_n(i)= \frac{y_n(i)^{1/T_{reg}}}{\sum \mathbf{y}_n(i)^{1/T_{reg}}}$}\\
			\STATE {$\Delta \mW_{l_{reg}}^{BP}=-\nabla (\mathbf{\hat{y}}_o(i)\cdot \log {\mathbf{\hat{y}}_n(i)})$}\\
   \STATE {$\mG = \gamma(i,j) \Delta \mW^{BP}_l(i,j)$}\\
   \STATE {$\mH = (1\!-\!\eta)\mR_l(j{-1}) \mG\!+\!\eta \Delta \mW_{l_{reg}}^{BP}(i,j)$}
   \STATE {$\mW_l(i,j)\!=\!\mW_l(i{-1},j)\!+\!\mH$}\\
\ENDIF
\ENDFOR
\ENDFOR
\end{algorithmic}
\end{algorithm}
represents the modification projector in our method. In Eq \ref{bp_awm}, we can easily observe that we modify weight direction adaptively by multiplying the BP gradient $\Delta \mW^{BP}_l(i,j)$ with our projector $\mR$ whose direction is varied according to the ratio $\beta$ of classes with similar feature distribution between different datasets and others.
\subsection{Regularization}
\label{regular}
There are also a few datasets where genuine audio is collected from quite different acoustic conditions compared with others. In this case, it is unreasonable to use the above method directly. As for these utterances, we introduce an extra regularization forcing the model to remember the previous inference distribution.\par
We first replicate the pre-trained model into two models with the same parameters, one is the teacher model and the other one is the student model. The parameter of the teacher model is frozen in the process of training on the new dataset and the parameter of the student model is fine-tuned. Like the operation in the LwF, we view the softmax output $\mathbf{y}_o$ from the teacher model as "soft labels" and use the loss function to slash the distinction between the "soft labels" $\mathbf{y}_o$ and the softmax output $\mathbf{y}_n$ of the student model, thus forcing the student model to remember the learned knowledge. 
The loss function, which is a modified cross-entropy loss, can be written as:
	\begin{equation}
	\small
	\mL_{reg}(\mathbf{\hat{y}}_o,\mathbf{\hat{y}}_n)\!=\!-\mathbf{\hat{y}}_o\cdot \log {\mathbf{\hat{y}}_n}
	\label{lwf_loss}
	\end{equation}
	\begin{equation}
	\small
	\hat{y}_o\!= \!\frac{y_o^{1/T_{reg}}}{\sum \mathbf{y}_o^{1/T_{reg}}}
	,\quad
	\hat{y}_n\!= \!\dfrac{y_n^{1/T_{reg}}}{\sum \mathbf{y}_n^{1/T_{reg}}}
	\end{equation}
where $T_{reg}$ is a constant hyperparameter. 
The $\mathbf{y_o}$, $\mathbf{y_n}$ are softmax outputs of teacher and student models, respectively; The $\mathbf{\hat{y}}$ is a normalized form of the $\mathbf{y}$; The $\hat{y}$ and $y$ are one item of $\mathbf{\hat{y}}$ and $\mathbf{y}$, respectively.
The weight modification of this regularization $\Delta \mW_{l_{reg}}^{BP}$ can be written as Eq \ref{lwf_w}.
	\begin{equation}
	\small
	\Delta \mW_{l_{reg}}^{BP}= \nabla \mL_{reg}
	\label{lwf_w}
	\end{equation}
	\par
\subsection{Regularized Adaptive Weight Modification}
\label{rawm}
In brief, our method RAWM is proposed for general continual learning conditions by modifying weight direction according to the ratio $\beta$ of classes with similar feature distribution between different datasets and others in batch data. By introducing a regularized restriction, our method eases the problem that a few data belonging to classes in the numerator of the Eq \ref{betaall} may have distinct feature distributions because they are collected from quite different conditions. 
Our method is inspired by fake audio detection where the ratio $\beta$ in the Eq \ref{betaall} can be written as:
\begin{small}
    \begin{equation}
    \label{betaonfake}
        \beta=\dfrac{N_g+1}{N_f+1}
    \end{equation}
\end{small}
in which $N_g$ and $N_f$ represent the number of genuine and fake audios in a batch, respectively. As for another research area, for example, speech emotion recognition including happy, sad, angry, and neutral, the essential ratio can be written as:
\begin{small}
    \begin{equation}
    \label{betaonemotion}
        \beta=\dfrac{N_{neu}+1}{N_{ang}+N_{hap}+N_{sad}+1}
    \end{equation}
\end{small}
because the neutral emotion has a relatively more similar feature distribution than others between different datasets. The $N_{neu}$, $N_{ang}$, $N_{sad}$, and $N_{hap}$ represent the number of neutral, angry, sad, and happy data in a batch, respectively.
\par
Considering a continual learning situation, the BP process of regularized adaptive weight modification can be written as Eq \ref{bp_rawm}.
\begin{small}
\begin{equation}
\begin{split}
\mW_l(i,j)\!=\!\mW_l(i{-1},j)\!+\!\mG \qquad j=1
\\
\mW_l(i,j)\!=\!\mW_l(i{-1},j)\!+\mH \qquad j>1
\\
\quad \mG = \gamma(i,j) \Delta \mW^{BP}_l(i,j) \qquad\qquad
\\
\quad \mH = (1\!-\!\eta)\mR_l(j{-1}) \mG\!+\!\eta \Delta \mW_{l_{reg}}^{BP}(i,j)
\end{split}
\label{bp_rawm}
\end{equation}
\end{small}
Compared with the Eq \ref{bp_awm}, our method introduces a regularization constraint to the adaptive weight modification. The importance of the regularization depends on the hyperparameter $\eta$ which is a coefficient measuring the attention degree of the knowledge acquired from old datasets.
\section{Experiments}
\label{results}
\begin{table*}[t]
\caption{(a) is the statistics of experimental datasets and (b) is the EER(\%) of our baseline on multiple evaluation sets.}
\begin{center}
\begin{subtable}[t]{0.54\linewidth}
\begin{center}
\caption{}
\label{datasetcount}
\resizebox{!}{1.cm}{
\begin{tabular}{c*{8}{c}}
\toprule[1.5pt]
\multirow{2.4}{*}{Dataset} & \multicolumn{2}{|c}{ASVSpoof2019LA\ ($\mathbf{S}$)} & \multicolumn{2}{|c}{ASVSpoof2015\ ($\mathbf{T}_1$)} & \multicolumn{2}{|c}{VCC2020\ ($\mathbf{T}_2$)} & \multicolumn{2}{|c}{In-the-Wild\ ($\mathbf{T}_3$)} \\
\cmidrule[0.2pt](){2-9}
& \multicolumn{1}{|c}{\#Real} & \#Fake & \multicolumn{1}{|c}{\#Real} & \#Fake & \multicolumn{1}{|c}{\#Real} & \#Fake & \multicolumn{1}{|c}{\#Real} & \#Fake\\
\midrule[1pt]
Train & \multicolumn{1}{|c}{2,580} & 22,800 & \multicolumn{1}{|c}{3,750} & 12,625 & \multicolumn{1}{|c}{1,330} & 3,060 & \multicolumn{1}{|c}{9,431} & 5,908\\
Dev & \multicolumn{1}{|c}{2,548} & 22,296 & \multicolumn{1}{|c}{3,497} & 49,875 & \multicolumn{1}{|c}{665} & 1,530 & \multicolumn{1}{|c}{4,715} & 2,954\\
Eval & \multicolumn{1}{|c}{7,355} & 63,882 & \multicolumn{1}{|c}{9,404} & 184,000 & \multicolumn{1}{|c}{665} & 1,530 & \multicolumn{1}{|c}{4,717} & 2,954\\
\bottomrule[1.5pt]
\end{tabular}}
\end{center}
\end{subtable}
\begin{subtable}[t]{0.45\linewidth}
\begin{center}
\caption{}
\label{train2019}
\resizebox{!}{0.5cm}{
\begin{tabular}{ccccc}
\toprule[1.pt]
\multicolumn{1}{c}{\bf Model}                &
\multicolumn{1}{c}{$\mathbf{S}$}                   &
\multicolumn{1}{c}{$\mathbf{T_1}$} &
\multicolumn{1}{c}{$\mathbf{T_2}$} &
\multicolumn{1}{c}{$\mathbf{T_3}$} 
\\ 
\midrule[0.5pt]
Baseline        &   $0.258$   &   $24.532$   & $46.503$ & $91.473$         \\
\bottomrule[1.pt]
\end{tabular}}
\end{center}
\end{subtable}
\end{center}
\end{table*}
\begin{table*}[t]
\caption{The EER(\%) on evaluation sets of our method with different $\eta$. (a), (b) and (c) are trained using the training set in order to $\mathbf{S} \rightarrow \mathbf{T_k}$ and are evaluated using the evaluation set on $\mathbf{S}$ and $\mathbf{T_k}$; (d) is trained using training set in order to $\mathbf{S} \rightarrow \mathbf{T_1} \rightarrow \mathbf{T_2} \rightarrow \mathbf{T_3}$ and is evaluated using evaluation sets.}
\begin{center}
\begin{subtable}[t]{0.22\linewidth}
\begin{center}
\caption{}
\label{ee2a}
\resizebox{!}{1.5cm}{
\begin{tabular}{ccc}
\toprule[1.pt]
\multicolumn{1}{c}{\bf $\eta$}                &
\multicolumn{1}{c}{$\mathbf{S}$}                   &
\multicolumn{1}{c}{$\mathbf{T_1}$}
\\ \midrule[0.5pt]
Baseline        &   $0.258$                 &   $24.532$            \\
\midrule[0.5pt]
$0.00$          &   $1.643$         &   $0.256$       \\
$0.20$          &   $1.424$         &   $0.431$       \\
$0.25$          &   $1.175$         &   $0.311$        \\
$0.50$          &   $0.878$         &   $0.257$    \\
$\mathbf{0.75}$ &   $\mathbf{0.666}$ &   $\mathbf{0.247}$             \\
$1.00$          &   $3.123$         &   $0.343$    \\
\bottomrule[1.pt]
\end{tabular}}
\end{center}
\end{subtable}
\begin{subtable}[t]{0.21\linewidth}
\begin{center}
\caption{}
\label{ee2b}
\resizebox{!}{1.5cm}{
\begin{tabular}{ccc}
\toprule[1.pt]
\multicolumn{1}{c}{\bf $\eta$}            &
\multicolumn{1}{c}{$\mathbf{S}$}               &
\multicolumn{1}{c}{$\mathbf{T_2}$}
\\ \midrule[0.5pt]
Baseline        &   $0.258$             &   $46.503$    \\
\midrule[0.5pt]
$0.00$          &   $1.413$         &   $3.845$    \\
$0.20$          &   $1.334$             &   $4.288$     \\
$0.25$          &   $1.275$             &   $3.994$     \\
$\mathbf{0.50}$          &   $\mathbf{1.237}$    &   $\mathbf{3.721}$     \\
$0.75$          &   $1.262$             &   $4.571$     \\
$1.00$          &   $4.234$             &   $4.566$     \\
\bottomrule[1.pt]
\end{tabular}} 
\end{center}
\end{subtable}
\begin{subtable}[t]{0.21\linewidth}
\begin{center}
\caption{}
\label{ee2c}
\resizebox{!}{1.5cm}{
\begin{tabular}{ccc}
\toprule[1.pt]
\multicolumn{1}{c}{\bf $\eta$}                &
\multicolumn{1}{c}{$\mathbf{S}$}                   &
\multicolumn{1}{c}{$\mathbf{T_3}$}
\\ 
\midrule[0.5pt]
Baseline        &   $0.258$                 &   $91.473$            \\
\midrule[0.5pt]
$0.00$          &   $6.126$                 &   $3.457$             \\
$0.20$          &   $5.490$                 &   $3.848$             \\
$0.25$          &   $4.975$                 &   $3.593$             \\
$\mathbf{0.50}$ &   $\mathbf{4.942}$        &   $\mathbf{3.249}$    \\
$0.75$          &   $4.482$                 &   $4.271$             \\
$1.00$          &   $4.453$                 &   $4.598$              \\
\bottomrule[1.pt]
\end{tabular}} 
\end{center}
\end{subtable}
\begin{subtable}[t]{0.34\linewidth}
\begin{center}
\caption{}
\label{evaleta4}
\resizebox{!}{1.5cm}{
\begin{tabular}{ccccc}
\toprule[1.pt]
\multicolumn{1}{c}{\bf $\eta$}     &
\multicolumn{1}{c}{$\mathbf{S}$}          &
\multicolumn{1}{c}{$\mathbf{T_1}$}          &
\multicolumn{1}{c}{$\mathbf{T_2}$}          &
\multicolumn{1}{c}{$\mathbf{T_3}$} \\ 
\midrule[0.6pt]
Baseline   &   $0.258$  &   $24.532$   &   $46.503$  &   $91.473$   \\
\midrule[0.5pt]
$0.00$         &   $1.845$  &   $1.127$    &   $3.916$   &   $3.410$   \\
$0.20$         &   $1.724$  &   $1.003$    &   $4.120$   &   $3.367$   \\
$0.25$         &   $1.699$  &   $0.945$    &   $4.017$   &   $3.529$   \\
$\mathbf{0.50}$         &   $\mathbf{1.508}$   &   $\mathbf{0.641}$  &   $\mathbf{3.850}$   &   $\mathbf{3.163}$  \\
$0.75$         &   $1.636$  &   $0.873$    &   $3.975$   &   $4.454$   \\
$1.00$         &   $2.714$  &   $1.621$    &   $3.875$   &   $4.325$   \\
\bottomrule[1.pt]
\end{tabular}
}
\end{center}
\end{subtable}
\label{evaleta2}
\end{center}
\end{table*}
\subsection{Datasets}
We conduct our experiments on four fake audio datasets, including the ASVspoof2019LA ($\mathbf{S}$), ASVspoof2015 ($\mathbf{T_1}$), VCC2020 ($\mathbf{T_2}$), and In-the-Wild ($\mathbf{T_3}$). The models are firstly trained using the training set of the $\mathbf{S}$ and are fine-tuned on the training sets of the other three datasets.
All of the experiments are evaluated using two or four evaluation sets in these datasets. The final model in the study refers to the model that was trained after the entire training process and then evaluated on each dataset.\par
\textbf{ASVspoof 2019 LA Dataset} \citep{todisco2019asvspoof} is the sub-challenge dataset (30 males and 37 females) containing three subsets: training, development, and evaluation. The training set and development share the same attack including four TTS and two VC algorithms. The bonafide audio is collected from the VCTK corpus \citep{veaux2017cstr}. The evaluation set contains totally different attacks. \par
\textbf{ASVspoof2015 dataset} \citep{wu2015asvspoof} is an open-source standard dataset of genuine and synthetic speech in the ASVspoof2015 challenge. The genuine speech was recorded from 106 speakers (45 males and 61 females) with no significant channel or background noise effects. The spoofing speech is generated using a variety of speech synthesis and voice conversion algorithms.\par
\textbf{VCC2020 dataset} \citep{zhao2020voice} is collected from Voice Conversion Challenge 2020. This dataset contains two subsets: a set of genuine audio provided by organizers and a set of fake audio provided by participating teams. Different from the previous three datasets, VCC2020 is a multilingual fake audio dataset, including English, Finnish, German and Mandarin.\par
\textbf{In-the-Wild dataset} \citep{muller2022does} contains a set of deep fake audio (and corresponding real audio) of 58 politicians and other public figures collected from publicly available sources, such as social networks and video streaming platforms. In total, 20.8 hours of genuine audio and 17.2 hours of fake audio were collected. On average, each speaker had 23 minutes of genuine audio and 18 minutes of fake audio.\par
We divide the genuine and fake audios of the VCC2020 dataset into four subsets. A quarter is used to build the evaluation set, a quarter to build the development set, and the rest to be used as the training set. The In-the-Wild dataset is divided in the same way as the VCC2020.
The detailed statistics of the datasets are presented in Table \ref{datasetcount}. The Equal Error Rate (EER), which is widely used for fake audio detection and speaker verification, is applied to evaluate the experimental performance.
\subsection{Experimental Setup}
\textbf{Fake Audio Detection Model}: We use the pre-trained model Wav2vec 2.0 \citep{2020wav2vec} as the feature extractor and the self-attention convolutional neural network (S-CNN) as the classifier. The parameters of Wav2vec 2.0 is loaded from the pre-train model XLSR-53 \citep{conneau2020unsupervised}. The classifier S-CNN contains three 1D-Convolution layers, one self-attention layer, and two full connection layers, according to the forward process. The input dimension of the first convolution layer is 256 and the hidden dimension of all convolution layers is 80. The kernel size and stride are set to 5 and 1, respectively. The hidden dimension of all full connection layers is 80 and the output dimension is 2.\par
\textbf{Training Details}: We fine-tune the model weights including the pre-trained model XLSR-53 and the classifier S-CNN. All of the parameters are trained by the Adam optimizer with a batch size of 2 and a learning rate $\gamma$ of 0.0001. The constant $\mathrm{m}$ and $T_{reg}$ in RAWM are set to 0.1 and 2, respectively. The $\alpha$ is initialized to 0.00001 for convolution layers, 0.0001 for the self-attention layer, and 0.1 for full connection layers. The norm in normalization of projector $\mP$ and $\mQ$ is the $\mL^2$ norm.
In addition, we present the results of training all datasets (Tain-on-All) that is considered to be the lower bound to all continual learning methods we mentioned \citep{DBLP:journals/nn/ParisiKPKW19}.
All results are (re)produced by us and averaged over 7 runs with standard deviations.
\subsection{Baseline}
We first train our model on the training set of the $\mathbf{S}$ dataset. Table \ref{train2019} shows the detection performance of our baseline on multiple evaluation sets which is very close to the state-of-the-art result \citep{DBLP:journals/tbbis/NautschWEKVTDSY21} in the same dataset. Although the model achieves promising performance on $\mathbf{S}$, its detection accuracy degrades significantly on other datasets. In addition, our baseline achieves the lowest cross-datasets EER on $\mathbf{T}_1$ dataset among three unseen datasets, which verifies that the detection model will have better performance when facing genuine audio with more similar feature distribution. 
Apart from that, the results with different training steps are presented in Table \ref{train2019_apd} in the appendix.
\begin{table*}[t]
\caption{The EER(\%) on evaluation sets of the ablation studies. (a), (b) and (c) are trained using the training set in order to $\mathbf{S} \rightarrow \mathbf{T_k}$ and are evaluated using the evaluation set on $\mathbf{S}$ and $\mathbf{T_k}$; (d) is trained in order to $\mathbf{S} \rightarrow \mathbf{T_1} \rightarrow \mathbf{T_2} \rightarrow \mathbf{T_3}$ and are evaluated using evaluation sets.}
\begin{center}
\begin{subtable}[t]{0.21\linewidth}
\small
\begin{center}
\caption{}
\label{evalab2a}
\resizebox{!}{0.8cm}{
\begin{tabular}{ccc}
\toprule[1.pt]
\multicolumn{1}{c}{\bf Method}                &
\multicolumn{1}{c}{$\mathbf{S}$}                   &
\multicolumn{1}{c}{$\mathbf{T_1}$}
\\ \midrule[0.5pt]
RAWM   &   $\mathbf{0.666}$        &   $\mathbf{0.247}$     \\
--REG  &   $1.643$    &   $0.256$             \\
--AWM &   $2.448$ &   $0.500$       \\
\bottomrule[1.pt]
\end{tabular}}
\end{center}
\end{subtable}
\small
\begin{subtable}[t]{0.21\linewidth}
\begin{center}
\caption{}
\label{evalab2b}
\resizebox{!}{0.8cm}{
\begin{tabular}{ccc}
\toprule[1.pt]
\multicolumn{1}{c}{\bf Method}            &
\multicolumn{1}{c}{$\mathbf{S}$}               &
\multicolumn{1}{c}{$\mathbf{T_2}$}
\\ \midrule[0.5pt]

RAWM &   $\mathbf{1.237}$        &   $\mathbf{3.721}$     \\
--REG             &   $1.413$                 &   $3.845$             \\
--AWM             &   $3.086$                 &   $5.432$             \\

\bottomrule[1.pt]
\end{tabular}} 
\end{center}
\end{subtable}
\begin{subtable}[t]{0.21\linewidth}
\begin{center}
\caption{}
\label{evalab2c}
\resizebox{!}{0.8cm}{
\begin{tabular}{ccc}
\toprule[1.pt]
\multicolumn{1}{c}{\bf Method}                &
\multicolumn{1}{c}{$\mathbf{S}$}                   &
\multicolumn{1}{c}{$\mathbf{T_3}$}
\\ \midrule[0.5pt]
RAWM              &   $\mathbf{4.942}$        &   $\mathbf{3.249}$     \\
--REG             &   $7.126$                 &   $3.357$             \\
--AWM             &   $8.130$                 &   $5.065$             \\				\bottomrule[1.pt]
\end{tabular}} 
\end{center}
\end{subtable}
\label{evalab2}
\begin{subtable}[t]{0.35\linewidth}
\begin{center}
\caption{}
\label{evalab4}
\resizebox{!}{0.8cm}{
\begin{tabular}{ccccc}
\toprule[1.pt]
\multicolumn{1}{c}{\bf Method}  &
\multicolumn{1}{c}{$\mathbf{S}$}       &
\multicolumn{1}{c}{$\mathbf{T_1}$}       &
\multicolumn{1}{c}{$\mathbf{T_2}$}       &
\multicolumn{1}{c}{$\mathbf{T_3}$}
\\ 
\midrule[0.5pt]
RAWM           &   $\mathbf{1.508}$   &   $\mathbf{0.641}$  &   $\mathbf{3.850}$   &   $\mathbf{3.163}$  \\
--REG         &   $1.845$  &   $1.127$    &   $3.916$   &   $3.410$   \\
--AWM         &   $4.083$  &   $2.167$    &   $6.480$   &   $5.472$   \\				
\bottomrule[1.pt]
\end{tabular}} 
\end{center}
\end{subtable}
\end{center}
\end{table*}
\begin{table*}
\caption{The EER(\%) of our method compared with various methods. (a), (b) and (c) are trained using the training set in order to $\mathbf{S} \rightarrow \mathbf{T_k}$ and are evaluated using the evaluation set on $\mathbf{S}$ and $\mathbf{T_k}$; (d) is trained using training set in order to $\mathbf{S} \rightarrow \mathbf{T_1} \rightarrow \mathbf{T_2} \rightarrow \mathbf{T_3}$ and is evaluated using evaluation sets.}
\begin{center}
\begin{subtable}[t]{0.23\linewidth}
\begin{center}
\caption{}
\label{em2a}
\resizebox{!}{1.4cm}{
\begin{tabular}{ccc}
\toprule[1.pt]
\multicolumn{1}{c}{\bf Method}                &
\multicolumn{1}{c}{$\mathbf{S}$}                   &
\multicolumn{1}{c}{$\mathbf{T_1}$}
\\ 
\midrule[0.5pt]
Baseline        &   $0.258$                 &   $24.532$            \\
Train-on-All    &   $0.406$                 &   $0.201$             \\
\midrule[0.5pt]
Fine-tune       &   $7.324$                 &   $0.510$             \\
EWC             &   $2.832$                 &   $0.570$             \\
OWM             &   $2.448$                 &   $0.540$             \\
LwF             &   $3.123$                 &   $0.343$             \\
DFWF            &   $1.849$                 &   $0.689$             \\
$\mathbf{RAWM(Ours)}$ &   $\mathbf{0.666}$        &   $\mathbf{0.247}$     \\
\bottomrule[1.pt]
\end{tabular}}
\end{center}
\end{subtable}
\begin{subtable}[t]{0.21\linewidth}
\begin{center}
\caption{}
\label{em2b}
\resizebox{!}{1.4cm}{
						\begin{tabular}{ccc}
							\toprule[1.pt]
							\multicolumn{1}{c}{\bf Method}            &
							\multicolumn{1}{c}{$\mathbf{S}$}               &
							\multicolumn{1}{c}{$\mathbf{T_2}$}
							\\ \midrule[0.5pt]
Baseline        &   $0.258$                 &   $46.503$            \\
Train-on-All       &   $0.965$                 &   $2.498$          \\
\midrule[0.5pt]
Fine-tune       &   $8.755$                 &   $5.647$             \\
EWC             &   $3.494$                 &   $6.289$             \\
OWM             &   $3.086$                 &   $6.432$             \\
LwF             &   $4.234$                 &   $4.566$             \\
DFWF            &   $1.874$                 &   $7.355$             \\
$\mathbf{RAWM(Ours)}$ &   $\mathbf{1.237}$        &   $\mathbf{3.721}$     \\
\bottomrule[1.pt]
\end{tabular}} 
\end{center}
\end{subtable}
\begin{subtable}[t]{0.21\linewidth}
\begin{center}
\caption{}
\label{em2c}
\resizebox{!}{1.4cm}{
\begin{tabular}{ccc}
\toprule[1.pt]
\multicolumn{1}{c}{\bf Method}     & \multicolumn{1}{c}{$\mathbf{S}$} & \multicolumn{1}{c}{$\mathbf{T_3}$} \\ \midrule[0.5pt]
Baseline         &             $0.258$              &              $91.473$      \\
Train-on-All     &             $2.740$              &              $2.160$       \\ 
\midrule[0.5pt]
Fine-tune        &             $20.976$             &              $4.978$       \\
EWC              &             $8.039$              &              $5.615$       \\
OWM              &             $8.130$              &              $5.065$       \\
LwF              &             $6.453$              &              $4.998$       \\
DFWF             &             $4.324$              &              $6.275$       \\
$\mathbf{RAWM(Ours)}$   & $\mathbf{4.942}$          &      $\mathbf{3.249}$      \\ \bottomrule[1.pt]
\end{tabular}} 
\end{center}
\end{subtable}
\begin{subtable}[t]{0.33\linewidth}
\begin{center}
\caption{}
\label{evalms4}
\resizebox{!}{1.4cm}{
						\begin{tabular}{ccccc}
							\toprule[1.pt]
							\multicolumn{1}{c}{\bf Method}  &
							\multicolumn{1}{c}{$\mathbf{S}$}       &
							\multicolumn{1}{c}{$\mathbf{T_1}$}       &
							\multicolumn{1}{c}{$\mathbf{T_2}$}       &
							\multicolumn{1}{c}{$\mathbf{T_3}$}
							\\ 
							\midrule[0.6pt]
Baseline   &   $0.258$  &   $24.532$   &   $46.503$  &   $91.473$   \\
Train-on-All  &   $1.324$  &   $0.561$   &   $3.579$  &   $2.008$  \\
\midrule[0.5pt]
Fine-tune   &   $7.068$  &   $2.841$    &   $5.674$   &   $4.543$   \\
EWC         &   $5.569$  &   $2.444$    &   $6.510$   &   $5.129$   \\
OWM         &   $4.083$  &   $2.167$    &   $6.480$   &   $5.472$   \\
LwF         &   $2.714$  &   $1.621$    &   $4.875$   &   $4.325$   \\
DFWF        &   $3.476$  &   $3.735$    &   $7.345$   &   $6.114$   \\
\bf{RAWM(Ours)} &   $\mathbf{1.508}$   &   $\mathbf{0.641}$  &   $\mathbf{3.850}$   &   $\mathbf{3.163}$  \\
\bottomrule[1.pt]
\end{tabular}} 
\end{center}
\end{subtable}
\label{evalms2}
\end{center}
\end{table*}
\subsection{Effects of the $\eta$ for our method}
\textbf{Sequence training between two datasets}: 
We start by performing some experiments to evaluate the effectiveness of $\eta$ in RAWM, which represents the attention degree to learned knowledge. In Table \ref{evaleta2}, we can easily observe that the RAWM achieves great performance on both old and new datasets, especially in the experiment on $\mathbf{S} \rightarrow \mathbf{T}_1$. By comparing the results of three cross-datasets, we observe that when the new and old datasets have similar feature distribution (Table \ref{ee2a}), there is an improvement in the performance of both acquiring new knowledge and overcoming forgetting with the increasing of $\eta$ ($\eta<1$); When the feature distribution of the new and old datasets is different (Table \ref{ee2b}, Table \ref{ee2c}), it is the model when $\eta=0.50$ that achieves the best result, which shows that the regularization we introduced is also of benefit to performance on both learning and overcoming forgetting.\par
\textbf{Sequence training on four datasets}:
We also present the results on multiple evaluation sets about different $\eta$ in Table \ref{evaleta4}. It can be observed that our method slashes performance degradation when training across datasets. The RAWM achieves the lowest EER among the results when $\eta=0.50$, which demonstrates that the same attention degree to both old and new datasets is the best choice for learning and overcoming forgetting. 
In addition, the results of $\mathbf{S}$, $\mathbf{T_1}$ and $\mathbf{T_2}$ show that the model with larger $\eta$ is more effective in overcoming forgetting.
\subsection{Ablation studies for our method}
\label{Ain2}
\textbf{Sequence training between two datasets}:
In this section, we compare our proposed method with adaptive weight modification without regularization (--REG) and orthogonal weight modification without regularization (--AWM). Table \ref{evalab2} presents their EER on three evaluation sets. We observe that RAWM achieves similar EER to --REG on the new dataset, both of them are superior significantly to --AWM, which shows that the adaptive weight modification has a significant positive impact on acquiring knowledge, while regularization impacts little. As for overcoming forgetting, when the feature distribution of the new and old datasets is similar (Table \ref{evalab2a}), the EER of the --REG on the old datasets is much lower than that of the --AWM and higher than that of the RAWM, which shows that the adaptive weight modification and regularization can significantly reduce the forgetting in this case. When the languages of the new and old datasets are different (Table \ref{evalab2b}), the EER of RAWM in the old datasets is similar to that of the --REG and much lower than that of the --AWM, which also proves that the adaptive weight modification has a significant positive impact on overcoming forgetting. When the feature distribution of the new and old datasets is quite different (Table \ref{evalab2c}), the EER of the --REG is similar to that of the --AWM and much higher than that of the RAWM, which shows that in this case, regularization is of great benefit to overcoming forgetting, while the effect of adaptive weight modification is not obvious.\par
\textbf{Sequence training on four datasets}:
In this section, we present the results of the ablation study on four evaluation sets in Table \ref{evalab4}. We observe that the EER of --REG to --AWM degrades more obviously than that of RAWM to --REG on all evaluation sets, which indicates that adaptive weight modification has a more obvious benefit in learning and overcoming forgetting than regularization for sequence training on multiple datasets. 
\subsection{Comparison of our method with other methods}
\textbf{Sequence training between two datasets}:
We compare our method with several methods in Table \ref{evalms2}. The EWC, LwF, and OWM as three mainstream continual learning methods achieve great success in many fields. The DFWF is the first continual learning method to overcome forgetting for fake audio detection. The results demonstrate that fine-tuning without modification (Fine-tune) forgets previous knowledge obviously. The forgetting of RAWM is one-tenth that of Fine-tune on Table \ref{em2a} and the EER on the new dataset of RAWM is also half that of Fine-tune. 
We also observe that the Fine-tune, EWC and OWM achieve similar performance in three experiments and the performance of LwF outperforms theirs on the new dataset. 
The DFWF is more effective in overcoming forgetting than the above methods, but its performance on the new dataset is inferior to others. Compared with others, our method achieves lower EER on both old and new datasets of all experiments, which demonstrates that both overcoming forgetting and learning could definitely benefit from our method when training across datasets, regardless of whether the datasets have similar feature distributions (Table \ref{em2a}, Table \ref{em2b}) or same languages (Table \ref{em2c}).\par
\textbf{Sequence training on four datasets}:
In addition, We compare our method with several methods for sequence training on four datasets in Table \ref{evalms4}. The results show that 
most methods achieve lower EERs than fine-tuning, and the best result for overcoming forgetting and learning is our proposed method, which indicates that the RAWM is superior to others for sequence training on both two and multiple datasets.
\subsection{The performance of the RAWM with a few samples}
\begin{table}[t]
\caption{The EER(\%) of few samples experiments. All experiments are first trained using the training set of $\mathbf{S}$ and then trained on 100 samples of the training set of $\mathbf{T}_1$. All experiments are evaluated using the evaluation set on $\mathbf{S}$ and $\mathbf{T}_1$.}
\begin{center}
\label{few-shot-100}
\resizebox{!}{1.5cm}{
\begin{tabular}{ccc}
				\toprule[1.pt]
				\multicolumn{1}{c}{\bf Method}                &
				\multicolumn{1}{c}{$\mathbf{S}$}                   &
				\multicolumn{1}{c}{$\mathbf{T}_1$}
				\\ \midrule[0.5pt]
				Baseline        &   $0.258$                 &   $24.532$            \\
				Train-on-All    &   $0.279$                 &   $0.291$
				\\\midrule[0.5pt]
				Fine-tune       &   $7.951$                 &   $0.617$             \\
				EWC             &   $2.972$                 &   $0.619$             \\
				OWM             &   $2.683$                 &   $0.617$             \\
				LwF             &   $3.198$                 &   $0.542$             \\
				DFWF            &   $1.975$                 &   $0.733$             \\
				$\mathbf{RAWM(Ours)}$ &   $\mathbf{0.923}$  &   $\mathbf{0.312}$     \\
\bottomrule[1.pt]
\end{tabular}}
\end{center}
\end{table}
We also present some results of the model when training on a few samples of new datasets. In our experiments, only 100 samples randomly selected from new datasets $\mathbf{T}_1$ were used for fine-tuning or continual learning. All models are first trained on the $\mathbf{S}$ datasets and then fine-tuned or continually learned on the $\mathbf{T}_1$ dataset. All models are trained on the new dataset within five steps. From the results, we can observe that our method RAWM also achieves the best performance on both old and new datasets and the learning performance is very close to the result of Train-on-All which is the the lower bound to all continual learning methods. By comparing the results in Table \ref{em2a} and Table \ref{few-shot-100}, we can easily find that reducing the number of samples has only a little damage to our method.
\subsection{The RAWM for speech emotion recognition}
\begin{table}[t]
\begin{small}
\caption{The Acc(\%) of various continual learning methods for 4-classes speech emotion recognition. All experiments are trained using the training set in order to $\mathbf{MSP}$-$\mathbf{Podcast}\!\rightarrow\!\mathbf{IEMOCAP}$ and are evaluated using the evaluation set on $\mathbf{MSP}$-$\mathbf{Podcast}$ and $\mathbf{IEMOCAP}$}
\begin{center}
\label{emotion}
\resizebox{!}{1.5cm}{
\begin{tabular}{ccc}
\toprule[1.pt]
\multicolumn{1}{c}{\bf Method}                          &
\multicolumn{1}{c}{$\mathbf{MSP}$-$\mathbf{Podcast}$}              &
\multicolumn{1}{c}{$\mathbf{IEMOCAP}$}
\\ \midrule[0.5pt]
Baseline        &   $54.446$                &   $30.043$            \\
Train-on-All    &   $54.396$                &   $57.262$             \\
\midrule[0.5pt]
Fine-tune       &   $24.094$                &   $50.379$             \\
EWC             &   $35.819$                &   $48.698$             \\
OWM             &   $32.267$                &   $48.162$             \\
LwF             &   $38.800$                &   $44.034$             \\
$\mathbf{RAWM(Ours)}$ &   $\mathbf{41.995}$        &   $\mathbf{54.229}$     \\
\bottomrule[1.pt]
\end{tabular}}
\end{center}
\end{small}
\end{table}
Our method is inspired by fake audio detection and it can be easily used in other related tasks. We take speech emotion recognition as an example to evaluate the performance of the RAWM in other fields. In this regard, the previous result shows that neutral emotion achieved the highest recognition accuracy across thirteen emotion datasets \citep{9747417}. So we infer that neutral speech has a more similar feature distribution than that of happy, sad, and angry, thus the ratio $\beta$ of our method can be written as Eq \ref{betaonemotion}.
Based on this observation, we conduct some experiments for speech emotion recognition. We choose four emotional classes, including neutral, happy, angry, and sad. The feature extractor and classifier are as same as that in fake audio detection. The results have been shown in Table \ref{emotion}. It could be easily observed that our method still achieves the highest accuracy on both datasets. 
The effect of our method in overcoming forgetting is most obvious and its learning performance is very close to the result of Train-on-All.
\subsection{The RAWM for image recognition}
\begin{table*}
\caption{The Accuracy(\%) on the CLEAR experiences. }
\label{clear}
\begin{center}
\resizebox{!}{2.5cm}{
\begin{tabular}{ccccccccccc}
\toprule[1.pt]
\multirow{2.6}{*}{\bf Continual Learning Methods} & 
\multicolumn{10}{c}{\bf Acc on the evaluation set of each experience}         \\
\cmidrule[0.5pt]{2-11}
~ & $\mathbf{Exp_1}$ & $\mathbf{Exp_2}$ & $\mathbf{Exp_3}$ & $\mathbf{Exp_4}$ & $\mathbf{Exp_5}$ & $\mathbf{Exp_6}$ & $\mathbf{Exp_7}$ & $\mathbf{Exp_8}$ & $\mathbf{Exp_9}$ & $\mathbf{Exp_{10}}$\\ 
\midrule[0.5pt]
Replay & $\mathbf{94.34}$ & $\mathbf{93.64}$ & $\mathbf{94.34}$ & $\mathbf{95.15}$ & $\mathbf{94.75}$ & $\mathbf{94.55}$ & $\mathbf{94.34}$ & $\mathbf{94.34}$ & $\mathbf{95.35}$ & $\mathbf{96.06}$\\
\midrule[0.5pt]
Fine-tune & $87.68$ & $90.00$ & $91.11$ & $91.82$ & $90.40$ & $89.90$ & $90.30$ & $90.61$ & $90.61$ & $93.33$\\
EWC & $84.04$ & $84.95$ & $85.86$ & $87.07$ & $85.66$ & $85.56$ & $86.97$ & $86.16$ & $85.76$ & $87.78$\\
LwF & $88.59$ & $88.89$ & $87.27$ & $90.51$ & $87.68$ & $87.78$ & $87.47$ & $87.47$ & $88.79$ & $88.48$\\
GDF & $91.11$ & $91.62$ & $88.38$ & $91.01$ & $88.79$ & $89.19$ & $90.20$ & $87.68$ & $90.10$ & $90.30$\\
CWR & $90.71$ & $91.72$ & $90.71$ & $91.52$ & $89.49$ & $90.91$ & $91.62$ & $90.71$ & $91.82$ & $93.74$\\
OWM & $91.62$ & $92.12$ & $\mathbf{91.82}$ & $93.64$ & $91.72$ & $\mathbf{92.42}$ & $92.22$ & $\mathbf{92.32}$ & $92.42$ & $95.05$\\
\midrule[0.5pt]
\textbf{RAWM (Ours)} & $\mathbf{92.12}$ & $\mathbf{92.53}$ & $91.41$ & $\mathbf{93.74}$ & $\mathbf{91.82}$ & $\mathbf{92.42}$ & $\mathbf{92.53}$ & $92.22$ & $\mathbf{92.53}$ & $\mathbf{95.25}$\\
\bottomrule[1.pt]
\end{tabular}
}
\end{center}
\end{table*}
We also have performed evaluation experiments on the image recognition domain in the CLEAR benchmark \citep{lin2021clear} to explore the broader applicability of our method. The CLEAR-10 benchmark for continual learning consists of 10 image recognition experiences, each comprising 11 classes such as camera, baseball, laptop, etc. To evaluate the effectiveness of our method, we selected several widely used continual learning algorithms, including the Replay, EWC, LwF, GDumbFinetune (GDF) \citep{DBLP:conf/eccv/PrabhuTD20}, CopyWeights with Re-init (CWR) \citep{DBLP:conf/corl/LomonacoM17}, and OWM methods. Table \ref{clear} presents the results of the comparative analysis. We treated the Replay method, which corresponds to the "Train-on-all" approach in our paper, as the upper bound of accuracy for all continual learning methods. 
The EWC and LwF methods have been introduced in our paper. In the GDF algorithm, we set the memory size to be the same as the number of training data in one bucket, and for CWR, the cwr layer was positioned as the final layer of the model. 
To extract features, we employed a pre-trained ResNet-50 \citep{he2016deep} as a feature extractor, producing 2048-dimensional feature vectors. A linear layer with input and output dimensions of 2048 and 11, respectively, was used as the downstream classifier. Our experiments were conducted with a batch size of 512, an initial learning rate of 1 (decayed by a factor of 0.1 after 60 epochs), and the SGD optimizer with a momentum of 0.9. The experimental results in Table \ref{clear} demonstrate that our proposed method, referred to as RAWM, consistently achieved the best performance across most tasks. In particular, in $\mathbf{Exp_3}$ and $\mathbf{Exp_8}$, the performance of our method closely approached the highest accuracy achieved among all the evaluated methods. 

\section{Conclusion}
In this work, we propose a continual learning algorithm to overcome catastrophic forgetting, called RAWM, that could adaptively modify the weight direction in process of training on new datasets. We also introduce a regularization to deal with the situation when old and new datasets are collected from quite different conditions. The experimental results demonstrate that our method outperforms four continual learning methods in learning and overcoming forgetting in scenarios of sequence training on both two and multiple datasets. The result shows that our method still achieves the best performance among the above methods when training on a few samples. Besides, our method is inspired by fake audio detection and the results show that it can be easily generalized to other fields, like speech emotion recognition. In addition, our method does not require previous data; thus it can be applied to most classification networks. We have yet to study how to make the model learn the weight direction gradually in the process of training on new datasets without any constraint, and exploring generalization to related tasks will form the focus of our future studies.

\section*{Acknowledgements}
This work is supported by the National Key Research and Development Program of China under Grant No.2020AAA0140003, the National Natural Science Foundation of China (NSFC) (No.61831022, No.U21B2010, No.62101553, No.61971419, No.62006223, No.62276259, No.62201572, No.62206278), Beijing Municipal Science and Technology Commission, Administrative Commission of Zhongguancun Science Park No.Z211100004821013, Open Research Projects of Zhejiang Lab (NO.2021KH0AB06).



\nocite{langley00}

\bibliography{example_paper}
\bibliographystyle{icml2023}

\newpage
\appendix
\onecolumn
\section{Appendix}

\begin{table}[htbp]
\caption{The EER(\%) on multiple evaluation sets. Model-1 to Model-6 are the models trained using the ASVspoof2019LA training set with increasing training steps.}
\label{train2019_apd}
\begin{center}
\resizebox{!}{1.6cm}{
\begin{tabular}{ccccc}
\toprule[1.pt]
\multirow{2.5}{*}{\bf Model} & 
\multicolumn{4}{c}{\bf Evaluation Sets}         \\
\cmidrule[0.5pt]{2-5}
~ & $\mathbf{S}$ & $\mathbf{T_1}$ & $\mathbf{T_2}$ & $\mathbf{T_3}$	\\ \midrule[0.5pt]
Model-1  &   $3.751$   &   $\mathbf{6.316}$   &   $\mathbf{7.670}$   &   $\mathbf{75.198}$  \\
Model-2  &   $2.975$   &   $8.517$   &   $10.000$   &   $78.477$  \\
Model-3  &   $1.794$   &   $9.988$   &   $26.165$   &   $85.436$  \\
\midrule[0.5pt]
Model-4  &   $\mathbf{0.258}$   &   $24.532$   &  $46.503$   &   $91.473$  \\
Model-5  &   $0.259$   &   $25.698$   &  $44.741$   &   $91.824$  \\
Model-6  &   $0.262$   &   $27.872$   &  $49.726$   &   $92.113$  \\
\bottomrule[1.pt]
\end{tabular}}
\end{center}
\end{table}

\end{document}